\documentclass[iop]{emulateapj}

\slugcomment{Accepted to ApJL}

\shorttitle{Nonmetastable Ammonia Masers in NGC~7538}
\shortauthors{Hoffman \& Kim}

\begin{document}

\title{New Maser Emission from Nonmetastable Ammonia in NGC~7538}

\author{Ian M.\ Hoffman}
\affil{St.\ Paul's School, Concord, NH 03301}
\email{ihoffman@sps.edu}

\and

\author{Stella Seojin Kim}
\affil{St.\ Paul's School, Concord, NH 03301}

\begin{abstract}
We present the first interferometric observations at 18.5~GHz of IRS~1 in NGC~7538.
These observations include images of the nonmetastable $^{14}$NH$_3$ $(9,6)$ masers with a synthesized beam of 2 arcseconds and images of the continuum emission with a synthesized beam of 150 milliarcseconds.
Of the maser emission, the previously known feature near $v_{\rm LSR} = -60\,{\rm km}\,{\rm s}^{-1}$ is spectrally resolved into at least two components and we observe several new maser emission features near $v_{\rm LSR} = -57\,{\rm km}\,{\rm s}^{-1}$.
The new maser emission near $-57\,{\rm km}\,{\rm s}^{-1}$ lies $250 \pm 90$~milliarcseconds northwest of the maser emission near $-60\,{\rm km}\,{\rm s}^{-1}$.
All of the masers are angularly unresolved indicating brightness temperatures $T_B > 2000\,{\rm K}$.
We are also able to conclusively associate the ammonia masers with the position of IRS~1.
The excitation of these rare ammonia masers is discussed in the context of the rich maser environment of IRS~1.
\end{abstract}


\section{Introduction}

The $(J,K) = (9,6)$ $^{14}$NH$_3$ maser was discovered by Madden et al.\ (1986) toward four of 17 Galactic star-forming regions surveyed: W51, W49, DR21(OH), and NGC~7538.
With the exception of confirmation observations shortly thereafter, the maser in NGC~7538 has not been observed since.
The maser in W51 was the subject of continued study ({\it e.g.}, Mauersberger et al.\ 1987; Wilson \& Henkel 1988), including the only interferometric observations of any these masers prior to the current study: a very long baseline experiment by Pratap et al.\ (1991).
Empirically, $^{14}$NH$_3$ (9,6) masers are known to exhibit intensity variability over several months (W49: Madden et al.\ 1986; W51: Wilson, Henkel, \& Johnston 1990).
In the case of the W51 masers, which are highly compact, Wilson et al.\ (1990) and Pratap et al.\ (1991) disagree as to whether the maser gain is saturated.
There is no model for the pump mechanism supplying the nonthermal population to the $(9,6)$ inversion transition for the stimulated amplification of the maser.
In general, nonmetastable $(J > K)$ transitions are not expected to support readily a population inversion, especially in the absence of masers in neighboring lines.
Nevertheless, Brown \& Cragg (1991) have studied ortho-ammonia ($K = 3n$, $n=0,1,...$) and found it possible to pump an isolated $(6,3)$ inversion, but they do not extend their model to the $(9,6)$ transition due to computational (though not physical) limitations.

The IRS~1 H~{\sc ii} region, assumed by Madden et al.\ to house the ammonia maser in the NGC~7538 complex, is particularly rich with maser species, both uncommon and ubiquitous (see Pratap et al.\ 1989; Hoffman et al.\ 2003; Galv\'an-Madrid et al.\ 2010 for a review).
Also, the contributions by an accretion disk, jet, and dust to the morphology of the continuum emission (see Sandell et al.\ 2009) and chemistry ({\it e.g.}, Gibb et al.\ 2001; Knez et al.\ 2009) are areas of ongoing study (see also Araya et al.\ 2007).
In a search for constraints on the complicated NGC~7538 region and on the rare masers it hosts, we have renewed the study of $(9,6)$ ammonia using the newly expanded frequency coverage of the EVLA.\footnote{The Very Large Array and National Radio Astronomy Observatory are facilities of the National Science Foundation operated under cooperative agreement by Associated Universities, Inc.}

\section{Observations and Results}

\subsection{`A' Configuration}

On 2008 October 15 we observed the 18.5-GHz line-free continuum emission from IRS~1 using the VLA correlator with 600-kHz bandwidth and the 16 antennas outfitted with EVLA receivers at the time.
The flux and phase calibrators used were 3C48 and 2322+509 with approximately 45 minutes of integration on the IRS~1 target.
The array was in `A' configuration resulting in a synthesized beam of approximately 150 milliarcseconds.
The primary beam of the VLA antennas at 18.5 GHz is 2.7 arcminutes.
The data were reduced using AIPS.\footnote{The Astronomical Image Processing System is documented at \url{http://www.aips.nrao.edu/}.}
The resulting image, shown in Figure~\ref{continuum}, has a background noise of $1\,{\rm mJy}\,{\rm beam}^{-1}$, consistent with instrumental expectations.
The total imaged flux density of IRS~1 at 18.5~GHz is within 10\% of 420~mJy.
The angular morphology of the emission is in good agreement with previous multi-wavelength studies (see, for example, Sandell et al.\ 2009).

\subsection{`DnC' Array}

On 2010 September 19 we observed the ammonia maser line emission (18499.390~MHz rest frequency) using the EVLA, employing 26 antennas and the WIDAR correlator.
As with the `A'-configuration observations of the continuum emission, the flux, phase, and bandpass calibrators used were 3C48 and 2322+509 with approximately 45 minutes of integration on the IRS~1 target.
The array was in `DnC' configuration resulting in a synthesized beam of 4.6 by 1.9 arcseconds at a position angle of 34 degrees.
The observations employed a 500-kHz bandwidth centered in frequency on $v_{\rm LSR}=-60\,{\rm km}\,{\rm s}^{-1}$ divided into 256 spectral channels resulting in a velocity resolution of $0.03\,{\rm km}\,{\rm s}^{-1}$ and a total velocity coverage of approximately $8\,{\rm km}\,{\rm s}^{-1}$.
The data were reduced using AIPS.
The {\it rms} background noise in a channel image is $17\,{\rm mJy}\,{\rm beam}^{-1}$, consistent with instrumental expectations.

The relative, deconvolved positions of the brightest image features are precise to within approximately 50~mas.
The position of the peak maser emission near $v_{\rm LSR} = -60\,{\rm km}\,{\rm s}^{-1}$ lies 250~mas $\pm$ 90~mas southeast of the peak maser emission near $v_{\rm LSR} = -57\,{\rm km}\,{\rm s}^{-1}$ and lies 150~mas $\pm$ 70~mas southeast of the location of the peak of continuum emission from IRS~1.
The image of the continuum emission from the VLA `A' observation has been registered in absolute position with the image of the continuum emission from the EVLA `DnC' observation in order to summarize these relative positions in Figure~\ref{continuum}.
The absolute position of the images are limited in precision to approximately 400~mas by interferometer phase solutions during data reduction.
The velocities in the EVLA spectrum corresponding to the different image positions are noted in Figure~\ref{line}.

\begin{deluxetable}{r@{.}l r@{.}l r@{.}l}
\tablecolumns{6}
\tablewidth{0pt}
\tablecaption{Fitted Spectral Properties of `SE' Maser Feature\label{table}}
\tablehead{
\multicolumn{2}{c}{$v_{\rm LSR}$} & \multicolumn{2}{c}{$I$} & \multicolumn{2}{c}{$\Delta{v}_{\rm FWHM}$} \\
\multicolumn{2}{c}{(${\rm km}\,{\rm s}^{-1}$)} & \multicolumn{2}{c}{(${\rm Jy}\,{\rm bm}^{-1}$)} & \multicolumn{2}{c}{(${\rm km}\,{\rm s}^{-1}$)}
}
\startdata
$-$60&21(2) & 0&80(4) & 0&49(3) \\
$-$59&79(3) & 0&52(6) & 0&48(3) \\ \tableline
\enddata
\tablecomments{The number in parentheses is the uncertainty in the final digit.}
\end{deluxetable}

The masers are not angularly resolved in the EVLA observations and have an upper limit on deconvolved size of approximately 2 arcseconds.
These sizes correspond to a lower limit on brightness temperature of $T_B > 2000\,{\rm K}$.
Assuming that the masers amplify the 18.5-GHz background continuum emission from IRS~1 following $T_B = T_{bg}e^{-\tau}$, we find their gains to be $\tau < -2$.
The spectra of the maser emission features are well sampled in the EVLA data and show an asymmetric shape to the emission near $v_{\rm LSR} = -60\,{\rm km}\,{\rm s}^{-1}$.
This shape is well fit as the sum of two coincident Gaussians who parameters are listed in Table~\ref{table} and whose curves are plotted with the data in Figure~\ref{line}.

The line emission at the edge of the observing band near $v_{\rm LSR} = -57\,{\rm km}\,{\rm s}^{-1}$ has not been detected previously (including the `NW' feature in Figure~\ref{line}).
The observations by Madden et al.\ in 1984 showed no line emission at these velocities and so this emission has increased in intensity by at least a factor of 50 in the last 27 years.
We have confirmed this new emission with subsequent observations using the Green Bank Telescope that will be described by us in a forthcoming paper.
This new emission also shows asymmetric line shapes indicative of several blended narrow maser lines.

\section{Discussion}

In concluding that the ammonia line emission from NGC~7538 is maser emission, several assumptions were made by Madden et al.\ (1986) since their arcminute-scale angular resolution did not constrain directly either the angular extent or the position of the line emission in the complex.
Namely, it was assumed that the masers were associated with IRS~1 and that they were no larger ($<2$ arcseconds) than the molecular clumps observed by other authors.
With the current data we are able to confirm both of these conjectures; definitively associating the line emission in position with the IRS~1 continuum emission and determining upper limits of approximately 2 arcseconds for the masering regions.

In addition, several maser components are spectrally resolved, having narrow line widths further suggestive of a maser explanation for the emission.
The fitted widths of these lines are approximately $0.5\,{\rm km}\,{\rm s}^{-1}$ which, if corresponding to Doppler shifts from thermal motion in the emitting gas, would indicate a kinetic gas temperature of approximately 90~K.
The (9,6) level is approximately 1100~K above the ground state.
The unphysically high brightness temperatures of the (9,6) ammonia emission lines in IRS~1 compared with their unphysically low temperatures calculated from Doppler-width assumptions strongly indicate a nonthermal, maser emission mechanism for the lines.

\begin{figure}
\epsscale{1.15}
\plotone{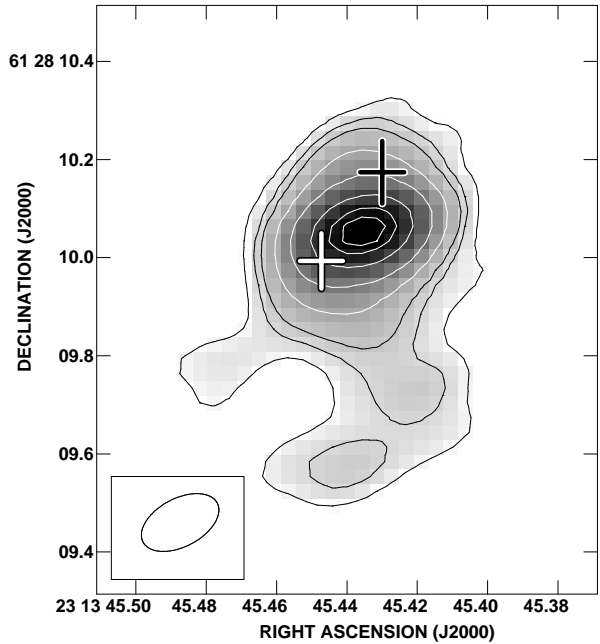}
\caption{
The positions of the $^{14}$NH$_3$ $(9,6)$ masers in NGC~7538 IRS~1 from the EVLA `DnC' configuration observations from 2010 September 19.
The crosses mark the positions of two of the maser features; the white cross for $v_{\rm LSR}=-60\,{\rm km}\,{\rm s}^{-1}$ (feature `SE' in Figure~\ref{line}) and the black cross for $v_{\rm LSR}=-57\,{\rm km}\,{\rm s}^{-1}$ (feature `NW' in Figure~\ref{line}).
The sizes of the crosses indicate the uncertainty in position of the masers.
The locations of CH$_3$OH, $^{15}$NH$_3$, and H$_2$CO masers in this region are discussed in \S3.
The image of the continuum emission from IRS~1 is the VLA `A' configuration image at 18.5~GHz from 2008 October 15.
The contour levels are $-$4, 4, 8, 12, 24, 48, 96, 144, and 176 times the image {\it rms} noise level of $1\,{\rm mJy}\,{\rm beam}^{-1}$.
The beam, plotted in the box at the lower left, is 171 by 98 milliarcsecond at a position angle of $-$62 degrees.
The absolute position uncertainties of the `DnC' and `A' observations are approximately 400~mas; for display in this figure, the continuum peak of the `A' image has been registered with the continuum peak of the `DnC' image from which the maser separations are known.
The higher precision of the relative positions of the masers from each other and from the continuum peak are described in \S2.
\label{continuum}}
\end{figure}

The pumping model of Brown \& Cragg (1991), which addresses explicitly the (6,3) transition and which is not physically inappropriate for the (9,6) transition, requires vibrational excitation of the ammonia molecules by infrared blackbody radiation from adjacent dust with a temperature of $\approx 300$~K.
If the kinetic gas temperature is also $\gtrsim 300$~K then the narrow, sub-Doppler lines widths are likely indicative of an unsaturated gain for the maser.
In contrast, the pumping process suggested by Brown \& Cragg (1991) readily saturates in terms of the availability of infrared photons.
These disparate results leave open the following two questions: (1) if the masers are unsaturated and if variability is a common hallmark of unmet saturation, then why does the intensity of the $v_{\rm LSR} = -60\,{\rm km}\,{\rm s}^{-1}$ (9,6) maser in NGC~7538 appear to have remained unvaried over 30 years? and
(2) if, on the other hand, the masers are saturated and the gain is linear across the line, then why are the line widths so narrow?
We note that a similar impasse has been reached in the interpretation of the observations of the (9,6) masers in W51 by Wilson et al.\ (1990) and Pratap et al.\ (1991).
In the absence of a well tested model for the excitation of the (9,6) maser, we compare the current results with other maser observations of IRS~1.

The angular resolution of the current data permit a discussion of the $^{14}$NH$_3$ (9,6) maser in the context of the other uncommon H$_2$CO, $^{15}$NH$_3$, and CH$_3$OH masers in this region.
The recent study of class-II (23.1-GHz) CH$_3$OH masers in IRS~1 by Galv\'an-Madrid et al.\ (2010) shows many similarities with nonmetastable ammonia.
Both of the main results of the current {\it Letter} are also seen for the methanol masers: (1) two spectral peaks of maser emission near $-60\,{\rm km}\,{\rm s}^{-1}$ and $-57\,{\rm km}\,{\rm s}^{-1}$ and (2) a $\sim$100~mas northwest-southeast angular separation between the peaks.
In seeking higher precision than the $\approx 100$-mas uncertainty in position registration between the CH$_3$OH masers and the archival image of the 15-GHz continuum emission from IRS~1 used for presentation by Galv\'an-Madrid et al., we have analyzed the archival data of the 23.1-GHz observations, including an image of the 23.1-GHz continuum emission.
Consistent within their quoted uncertainties, we find that the northern and southern CH$_3$OH masers lie on opposite sides of the peak of the 23.1-GHz continuum emission.
The location of the more redshifted maser $\approx 40$~mas to the north of the continuum peak and of the blueshifted maser $\approx 110$~mas to the south is also seen qualitatively in the current $^{14}$NH$_3$ distribution, indicating that the two species lie in the same kinematic volumes of cloud and outflow, though not necessarily in the same clumps of physical conditions.
Unlike $^{14}$NH$_3$, the 23.1-GHz CH$_3$OH masers show broad ($2.5\,{\rm km}\,{\rm s}^{-1}$) line widths and little, if any, intensity variability over 25 years, likely indicative of saturated maser gain.

The IRS~1 region also exhibits emission from $^{15}$NH$_3$, a molecular species that is comparable chemically to the $^{14}$NH$_3$ presented currently, though likely not similar in excitation.
Gaume et al.\ (1991) find a distribution of metastable $^{15}$NH$_3$ (3,3) 22.8-GHz masers having many similarities with the current presentation of $^{14}$NH$_3$ (9,6) masers.
In particular, of the numerous $^{15}$NH$_3$ masers, there are two distinct groupings whose velocities and NW-SE position displacement agree well with our Figure~\ref{continuum}.
Two other groupings of $^{15}$NH$_3$ masers, for which there is no correlated $^{14}$NH$_3$ emission, extend to redder and bluer velocities from which Gaume et al.\ suggest a large scale SW-NE velocity pattern of which the $^{14}$NH$_3$ velocity distribution presented in our Figure~\ref{continuum} is but a small part.
We are undertaking wider bandwidth observations of the $^{14}$NH$_3$ masers, as well as interferometric observations of the $^{15}$NH$_3$ masers in a modern epoch, that will permit determination of any additional correlations among the new $^{14}$NH$_3$ masers at $v_{\rm LSR} \geq -57\,{\rm km}\,{\rm s}^{-1}$ and the $^{15}$NH$_3$ masers which, 22 years ago, extended over the velocity range $-62\,{\rm km}\,{\rm s}^{-1} < v_{\rm LSR} < -52\,{\rm km}\,{\rm s}^{-1}$.
Gaume et al.\ note that the position registration of the $^{15}$NH$_3$ masers with respect to well-modeled images of the continuum emission from IRS~1 is plagued by the same absolute position uncertainties we have discussed here for other species.
Nevertheless, the relative location of the ammonia masers and continuum emission are well known in the data of Gaume et al.\ and the current data, respectively, and are in agreement with each other.

There are also 4.8-GHz H$_2$CO masers at $v_{\rm LSR} = -60.1\,{\rm km}\,{\rm s}^{-1}$ and $v_{\rm LSR} = -57.9\,{\rm km}\,{\rm s}^{-1}$ in IRS~1.
The two interferometric data sets sensitive to both the continuum and line emission (Rots et al.\ 1981; Hoffman et al.\ 2003) find both of the H$_2$CO masers to lie west of the 6-cm continuum peak and find the $-60.1\,{\rm km}\,{\rm s}^{-1}$ feature to be the westernmost feature.
The positions of the formaldehyde velocity components stand in contrast to the moderate position and velocity correlation among CH$_3$OH, $^{14}$NH$_3$, and $^{15}$NH$_3$ in which the $-60\,{\rm km}\,{\rm s}^{-1}$ feature lies to the east of the continuum peak and of the redshifted, $-57\,{\rm km}\,{\rm s}^{-1}$ maser.
The H$_2$CO masers are variable: there is long-term intensity variability in the H$_2$CO masers which may be correlated with the appearance of the new $^{14}$NH$_3$ masers reported in this {\it Letter}.

\begin{figure*}
\includegraphics[angle=270,scale=0.75]{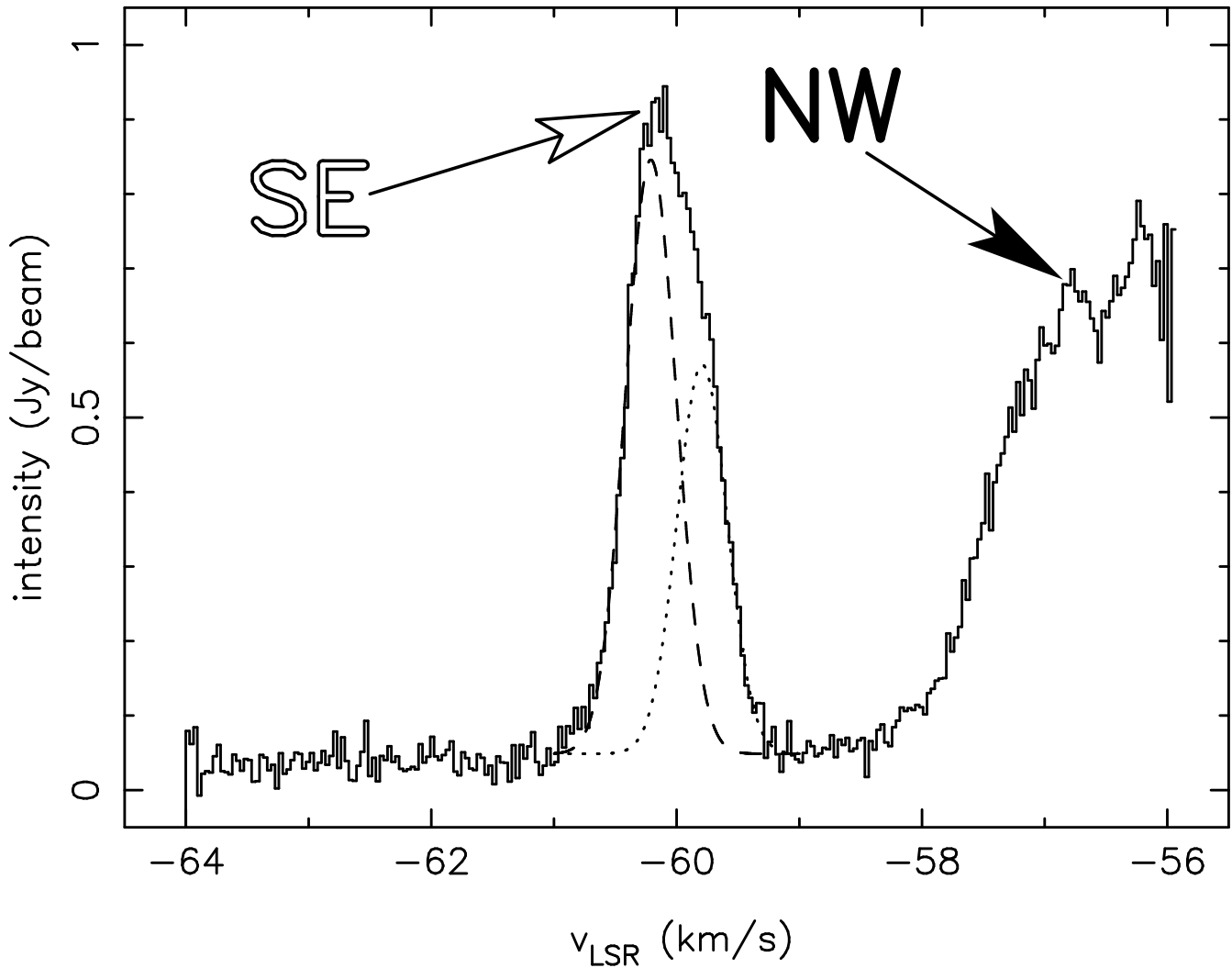}
\caption{
Spectrum of the $^{14}$NH$_3$ $(9,6)$ masers in NGC~7538 IRS~1 from the EVLA `DnC' configuration observations from 2010 September 19 (solid histogram) after subtraction in the $u,v$ plane of the contribution from continuum emission.
The `SE' and `NW' labels correspond to the deconvolved locations of the emission in the image plane of Figure~\ref{continuum}.
This single spectrum includes all of the maser emission since all of the emission is convolved within the synthesized beam on the sky of size 4.6 by 1.9 arcseconds at a position angle of 34 degrees.
The dotted and dashed lines are the two best-fit Gaussians whose sum is the asymmetric line profile near $v_{\rm LSR}=-60\,{\rm km}\,{\rm s}^{-1}$, as summarized in Table~\ref{table} and described in \S2.
\label{line}}
\end{figure*}

Common temporal variability of various maser species may provide insight into their various excitation conditions.
For example, two maser species may exist at different densities or velocities or positions, though both may be dependent on the same infrared source for radiative excitation and therefore exhibit common variability.
Or, the relatively slow passage of a bulk material disturbance through the maser environments will affect only those masers sufficiently sensitive to changes in density or collisions when the disturbance arrives at their location.
We examine possible excitation scenarios in IRS~1 in order to explain two differing aspects of the $^{14}$NH$_3$ masers reported here: (1) the unchanged intensity of the component near $v_{\rm LSR} = -60\,{\rm km}\,{\rm s}^{-1}$ and (2) the $>50$-fold increase in intensity of the component near $v_{\rm LSR} = -57\,{\rm km}\,{\rm s}^{-1}$.

Neither the CH$_3$OH nor $^{15}$NH$_3$ masers discussed above have exhibited any intensity variability in either the $-60\,{\rm km}\,{\rm s}^{-1}$ or $-57\,{\rm km}\,{\rm s}^{-1}$ velocity components since their discovery approximately 30 years ago (Wilson et al.\ 1984; Johnston et al.\ 1989).
Since class-II methanol masers are thought to depend on excitation from infrared radiation, the current $^{14}$NH$_3$ variability results are not suggestive of a common radiative excitation for ammonia and methanol.
Hoffman et al.\ (2003) note common, long-term variability of H$_2$CO and H$_2$O masers, further described by Lekht et al.\ (2003, 2004; see also Hoffman et al.\ 2007).
Araya et al.\ (2007) have continued to monitor the variability of the two H$_2$CO velocity components, suggesting possible successive excitation by a moving disturbance or jet similar to that described by Araya et al.\ (2009) to explain the behavior of CH$_3$OH masers in DR21(OH).
It is possible that the $^{14}$NH$_3$ maser near $v_{\rm LSR} = -57\,{\rm km}\,{\rm s}^{-1}$ has been excited sometime in the past 27 years by the arrival of a bulk material disturbance.
Such a disturbance, if responsible for all of the long-term variability in H$_2$CO, NH$_3$, and H$_2$O, would need to be an isotropic flow or jet directed both north and south from the center of IRS~1 since the H$_2$CO maser and the variable nonmetastable ammonia maser lie on opposite sides of the peak of continuum emission.
Ongoing monitoring observations are essential for determining the presence of an excitation disturbance traversing the region.
Since Brown \& Cragg (1991) find the pumping of the (6,3) line, and likely also the (9,6) line, to require both radiative and collisional excitation, the existence of the rare ammonia, formaldehyde, and methanol masers in IRS~1 may be indicative of a relatively uncommon superposition of infrared radiation and material outflow.

The radio continuum morphology of IRS~1 is extended north-south and has been successfully modeled as a bipolar outflow with the southern half more blueshifted than the north (for example, Sandell et al.\ 2009).
The various maser species discussed above have been speculated inconclusively to include some or all of a jet, a shock, microwave free-free emission, infrared dust emission, and molecular clumps for their excitations.
Since relative positions within a data set are generally more precise than comparisons of absolute positions between data sets, we have restricted our discussion in this {\it Letter} to the relative positions of the NH$_3$, H$_2$CO, and CH$_3$OH masers and C- and K-band images of the continuum emission.
Nevertheless, in addition to the broad correlations we have noted, other authors have suggested several specific associations among individual masers (additionally H$_2$O and OH) and among thermal and continuum features that are supported by precise absolute positions ({\it e.g.}, Hoffman et al.\ 2003; Galv\'an-Madrid et al.\ 2010).
We plan to explore possible associations and interrelated physical conditions in a forthcoming paper.

\section{Conclusion}

We report the discovery of new maser emission from the nonmetastable $(J,K) = (9,6)$ inversion transition of $^{14}$NH$_3$ in NGC~7538 IRS~1.
Both the previously known feature near $v_{\rm LSR} = -60\,{\rm km}\,{\rm s}^{-1}$ and the new masers near $-57\,{\rm km}\,{\rm s}^{-1}$ are imaged interferometrically for the first time, showing different positions in IRS~1 in correlation with some other maser species.
The brightness temperatures ($T_B > 2000$~K) and line widths ($0.5\,{\rm km}\,{\rm s}^{-1}$) of the features demonstrate conclusively the nonthermal nature of the emission.
We plan future EVLA observations of these masers with improved angular resolution in order to constrain their excitation mechanism.

\acknowledgments

This work is supported by the Thomas Penrose Bennett Prize Fund and the Lovejoy Science Fund of St.\ Paul's School.
We are grateful to the NRAO staff in Socorro for flexible scheduling and guidance during the EVLA transition.
We are indebted to an anonymous referee for comments that significantly improved the manuscript.

{\it Facilities:} \facility{VLA ()}, \facility{EVLA ()}


\begin{thebibliography}{}
\bibitem[Araya et al.\ (2007)]{ara07} Araya, E., Hofner, P., Goss, W.\ M., Linz, H., Kurtz, S., Olmi, L.\ 2007, \apjs, 170, 152
\bibitem[Araya et al.\ (2009)]{ara09} Araya, E., Kurtz, S., Hofner, P., Linz, H.\ 2009, \apj, 698, 1321
\bibitem[Brown \& Cragg (1991)]{bro91} Brown, R.\ D., Cragg, D.\ M.\ 1991, \apj, 378, 445
\bibitem[Galv\'an-Madrid et al.\ (2010)]{gal10} Galv\'an-Madrid, R., Montes, G., Ramírez, E.\ A., Kurtz, S., Araya, E., Hofner, P.\ 2010, \apj, 713, 423
\bibitem[Gaume et al.\ (1991)]{gau91} Gaume, R.\ A., Johnston, K.\ J., Nguyen, H.\ A., Wilson, T.\ L., Dickel, H.\ R., Goss, W.\ M., Wright, M.\ C.\ H.\ 1991, \apj, 376, 608
\bibitem[Gibb et al.\ (2001)]{gib01} Gibb, E.\ L., Whittet, D.\ C.\ B., Chiar, J.\ E.\ 2001, \apj, 558, 702
\bibitem[Hoffman et al.\ (2003)]{H03} Hoffman, I.\ M., Goss, W.\ M., Palmer, P., Richards, A.\ M.\ S.\ 2003, \apj, 598, 1061
\bibitem[Hoffman et al.\ (2007)]{H07} Hoffman, I.\ M., Goss, W.\ M., Palmer, P.\ 2007, \apj, 654, 971
\bibitem[Johnston et al.\ (1989)]{joh89} Johnston, K.\ J., Stolovy, S.\ R., Wilson, T.\ L., Henkel, C., Mauersberger, R.\ 1989, \apjl, 343, L41
\bibitem[Knez et al.\ (2009)]{kne09} Knez, C., Lacy, J.\ H., Evans, N.\ J., van Dishoeck, E.\ F., Richter, M.\ J.\ 2009, \apj, 696, 471
\bibitem[Lekht et al.\ (2003)]{lek03} Lekht, E.\ E., Munitsyn, V.\ A., Tolmachev, A.\ M.\ 2003 Astron.\ Rep.\ 47, 838
\bibitem[Lekht et al.\ (2004)]{lek04} Lekht, E.\ E., Munitsyn, V.\ A., Tolmachev, A.\ M.\ 2004 Astron.\ Rep.\ 48, 200
\bibitem[Madden et al.\ (1986)]{mad86} Madden, S.\ C., Irvine, W.\ M., Matthews, H.\ E., Brown, R.\ D., \& Godfrey, P.\ D.\ 1986, \apjl, 300, L79
\bibitem[Mauersberger et al.\ (1987)]{mau87} Mauersberger, R., Henkel, C., Wilson, T.\ L.\ 1987, \aap, 173, 352
\bibitem[Pratap et al.\ (1989)]{pra89} Pratap, P., Batrla, W., Snyder, L.\ E.\ 1989, \apj, 341, 832
\bibitem[Pratap et al.\ (1991)]{pra91} Pratap, P., Menten, K.\ M., Reid, M.\ J., Moran, J.\ M., Walmsley, C.\ M.\ 1991, \apjl, 373, L13
\bibitem[Rots et al.\ (1981)]{rot81} Rots, A.\ H., Dickel, H.\ R., Forster, J.\ R., Goss, W.\ M.\ 1981, \apjl, 245, L15
\bibitem[Sandell et al.\ (2009)]{san09} Sandell, G., Goss, W.\ M., Wright, M., \& Corder, S.\ 2009, \apjl, 699, L31
\bibitem[Wilson et al.\ (1984)]{wil84} Wilson, T.\ L., Walmsley, C.\ M., Snyder, L.\ E., Jewell, P.\ R.\ 1984, \aap, 134, L7
\bibitem[Wilson \& Henkel (1988)]{wil88} Wilson, T.\ L., Henkel, C.\ 1988, \aap, 206, L26
\bibitem[Wilson et al.\ (1990)]{wil90} Wilson, T.\ L., Henkel, C., Johnston, K.\ J.\ 1990, \aap, 229, L1
\end{thebibliography}
\end{document}